\documentclass{ws-ijmpa}

\begin{document}

\markboth{D. J. Summers}
{6D Ionization Muon Cooling with Tabletop Rings}

\catchline{}{}{}{}{}
%
%%%%%%%%%%%%%%%%%%%%%%%%%%%%%%%%%%%%%%%%%%%%%%%%%%%%%%%%%%%%%%%%%%%%

\title{6D IONIZATION MUON COOLING WITH TABLETOP RINGS}

\author{\footnotesize D. J. SUMMERS,\footnote{summers@phy.olemiss.edu
\quad Supported by DE--FG02--91ER40622.} \ \ S. B. BRACKER, \ L. M. CREMALDI, 
\ and R. GODANG}

\address{Department of Physics and Astronomy, 
University of Mississippi--Oxford\\
University, MS 38677, USA}

\author{\footnotesize D. B. CLINE and A. A. GARREN}

\address{Department of Physics, 
University of California--Los Angeles\\
Los Angeles, CA 90095, USA}

\author{\footnotesize G. G. HANSON and A. KLIER}

\address{Department of Physics, 
University of California--Riverside\\
Riverside, CA 92521, USA}

\author{\footnotesize S. A. KAHN, H. G. KIRK, and R. B. PALMER}

\address{Brookhaven National Laboratory\\
Upton, NY 11973, USA}

\maketitle

\pub{Received (Day Month Year)}{}
%{Revised (Day Month Year)}

\begin{abstract}
Progress on six dimensional ionization muon cooling with relatively small rings
of magnets is described. Lattices being explored include scaling 
sector cyclotrons
with edge focusing and strong focusing, fixed field alternating gradient
(FFAG) rings. Ionization cooling is provided by high pressure hydrogen gas
which removes both transverse and longitudinal momentum. Lost longitudinal
momentum is replaced using radio frequency (RF) cavities, giving a 
net transverse
emittance reduction. The longer path length in the hydrogen of higher momentum
muons decreases longitudinal emittance at the expense of transverse emittance.
Thus emittance exchange allows these rings to cool in all six dimensions and
not just transversely. 
Alternatively, if the RF is located after the ring, it may be possible to cool
the muons by stopping them as they spiral adiabatically into a
central swarm. As $p \to 0$, $\Delta{p} \to 0$.
The resulting cooled muons can lead to an intense muon
beam which could be a source for neutrino factories or muon colliders.

\end{abstract}

\section{Introduction}	%) A SECTION HEADING

Muons at rest have a  2.2 $\mu$s lifetime; cooling an ensemble of muons 
must be completed faster than that.
Ionization cooling can help.$^1$ Random muon motion is removed by passage
through a low Z material, such as hydrogen, and coherent motion is added with
RF acceleration. Designs to cool muons in six dimensions using linear
helical channels$^{\,2}$  at 100 MeV kinetic energies and using frictional
cooling$^{\,3}$ at 1 keV kinetic energies are under investigation. A number of
muon cooling rings have been simulated at various levels.$^4$ 
In a ring structure,
the same magnets
and RF cavities may be reused each time a muon orbits. Transverse cooling can
naturally be exchanged for longitudinal cooling by allowing higher
momentum muons to pass through more material. Thus rings can cool in all six
dimensions.  The rings reported here are the smallest to date, and are
basically radial sector cyclotrons.

Small emittance bunches of cold muons are useful for a neutrino factory$^{\,5}$
and
are required for a muon collider.$^6$ 
At a neutrino factory, accelerated muons are stored in a
racetrack to produce neutrino beams
($\mu^- \to e^- \, {\overline{\nu}}_e \, \nu_{\mu}$ \, and \,
$\mu^+ \to e^+ \, \nu_e \, {\overline{\nu}}_{\mu}$). Neutrino oscillations
have been observed$^{\,7}$ and need more study. Further
exploration at a neutrino factory could reveal CP
violation in the lepton sector,$^{8}$ and will be  particularly useful if the
$\nu_e$ to $\nu_{\tau}$ coupling, $\theta_{13}$, is small.$^9$
A muon collider can do s-channel scans to split
the $H^0$ and $A^0$ Higgs doublet.$^{10}$
Above the ILC's 800 GeV there are a large array of supersymmetric particles
that might be produced$^{\,11}$ and,
if large extra dimensions exist,
so could mini black holes.$^{12}$
Note that the energy resolution of a 4 TeV muon
collider is not smeared by beamstrahlung.

\begin{figure}[b]
\begin{minipage}{0.45\textwidth}
%\centerline{\psfig{file=blackboard_ring_new.eps,width=6cm}}
\centerline{\psfig{file=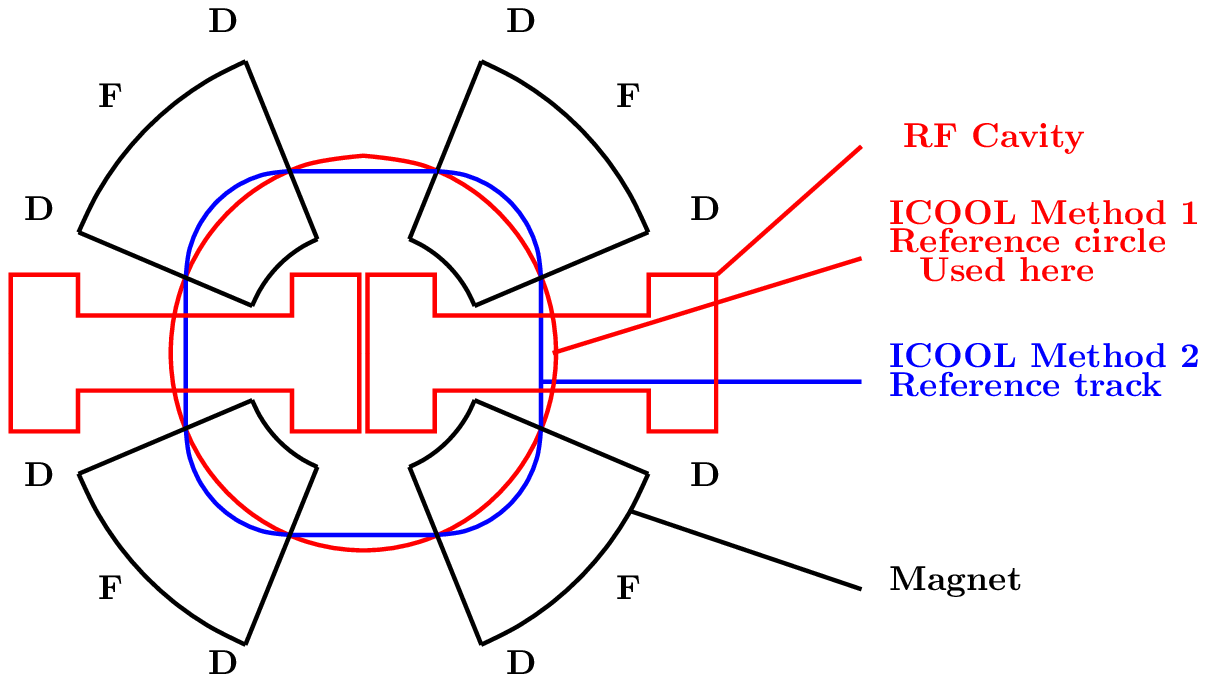,width=6cm}}
%\vspace*{8pt}
\caption{ A way of fitting a pair of large 201 MHz RF cavities into a small 
four sector cyclotron.  The cavities fill the center of the cyclotron.}
%The four hashed areas around the periphery are return yokes for the magnets.} 
\end{minipage} \hfill
\begin{minipage}{0.45\textwidth}
\centerline{\psfig{file=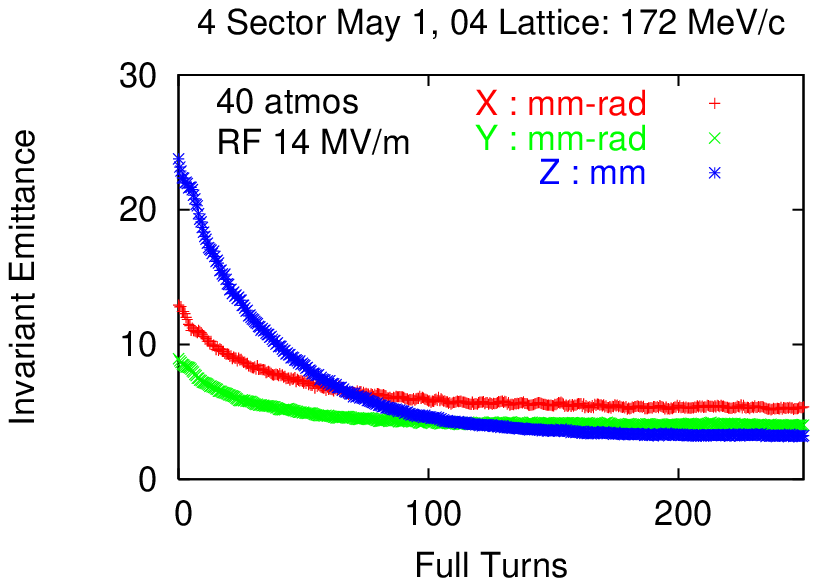,width=6cm}}
%\vspace*{8pt}
\caption{X, Y, and Z emittance versus orbits for the low field ring
with four sectors that is described in Table 1.}
\end{minipage}
\end{figure}

\section{Hydrogen Gas Filled Sector Cyclotrons with Internal RF}

A low field ring is being designed using ICOOL$^{\,13}$ 
for optimization and SYNCH$^{\,14}$ for lattices.
It could be built to demonstrate six dimensional muon cooling. Relatively
constant $\beta$ functions allow a continuous placement of the energy absorber.
Hydrogen gas cools best and also prevents breakdown in the 201 MHz RF
cavities.$^2$ Skew quadrupole magnets mix vertical and horizontal
betatron oscillations. The parameters for the ring appear in Table 1. 
Simulated 6D cooling is shown in Fig. 2.

The equation describing transverse cooling (with energies in GeV) is:
  \begin{equation}
\frac{d\epsilon_n}{ds}\ =\
-\frac{1}{\beta^2} \frac{dE_{\mu}}{ds}\ \frac{\epsilon_n}{E_{\mu}}\ +
\ \frac{1}{\beta^3} \frac{\beta_{\perp} (0.014)^2}{2\ E_{\mu}m_{\mu}\ L_R},
\label{eq1}
  \end{equation}
where $\beta =$ v/c,
$\epsilon_n$ is the normalized emittance, $\beta_{\perp}$ is the
betatron function (focal length)
at the absorber, $dE_{\mu}/ds$ is the energy loss, and $L_R$
is the radiation length of the material.  The first term in this equation is
the cooling term, and the second is the heating term due to multiple
scattering. This heating term is minimized if $\beta_{\perp}$ is small
(strong-focusing) and $L_R$ is large (a low-Z absorber).
The equilibrium emittance is  achieved when the heating and cooling terms 
balance. 

A higher field ring with smaller values of $\beta_{\perp}$ can give 
a higher cooling merit factor.
Simulations show that high RF gradients are required
as noted in Table 1.

A scaling fixed field, alternating gradient (FFAG) ring$^{\,15}$ can 
allow the use of lower RF gradients and magnetic fields.  
For the FFAG ring in Table 1, a 
focusing parameter, $n = -(r/B) (dB/dr) = -0.6$, was used. 
Each sector has a focusing-defocusing-focusing (FDF) geometry with 
three magnets.
There are no open slots 
between sectors, so the RF cavities would have to placed
within the magnets. 

\begin{table}[t]
\tbl{Scaling sector cyclotron parameters. The cooling merit factors 
include transmission
loss; the X, Y, and Z cooling factors do not. 
The low field ring is designed to test 6D
cooling.} 
{\begin{tabular}{@{}lccc@{}} \toprule
              & Low Field Ring & High Field Ring & FFAG Ring \\ \colrule
Dipole Field       & 1.8 T     & 5.2 T & 2.6 T\\
Magnetic Elements  & hard edge & hard edge & hard edge \\
Sectors            & 4         & 6  & 12 \\
Focusing           & edge      & edge & alternating gradient \\
$\beta_x$ Range    & 38 $\to$ 92 cm  & 26 $\to$ 36 cm  & 44 $\to$ 65 cm  \\
$\beta_y$ Range    & 54 $\to$ 66 cm  & 30 $\to$ 32 cm  & 26 $\to$ 42 cm \\
p(central orbit)   & 172 MeV/c & 250 MeV/c & 250 MeV/c \\
Hydrogen Pressure  & 40 Atm. @ 300$^{\,0}$ K & 100 Atm. @ 300$^{\,0}$ K & 
                                                   100 Atm. @ 300$^{\,0}$ K\\
Peak RF Gradient   & 14 MV/m   & 45 MV/m & 8 MV/m \\
Total RF Length    & 1.6 m     & 0.8 m & 3.6 m  \\
Ring Circumference & 3.81 m    & 1.95 m & 6.0 m \\
X Aperture      & $\pm$20 cm  & $\pm$25 cm  & $\pm$25 cm \\
Y Aperture      & $\pm$15 cm  & $\pm$15 cm  & $\pm$15 cm  \\
$p_z$ Acceptance & $\pm$10 MeV/c & $\pm$10 MeV/c & $\pm$10 MeV/c \\
Orbits             & 100       & 250 & 40 \\
X, Y, Z Cooling Factors  & 2.4, \ 2.3, \ 7.4 & 2, \ 13, \ 20 & 16, \ 15, \ 2 \\
Muon Decay Included & no       & yes & yes \\
Cooling Merit Factor  & 20        & 400 & 120\\
\botrule 
\end{tabular}}
\end{table}

\begin{figure}[b]
\vspace*{-10mm}
\begin{minipage}{0.39\textwidth}
\centerline{\psfig{file=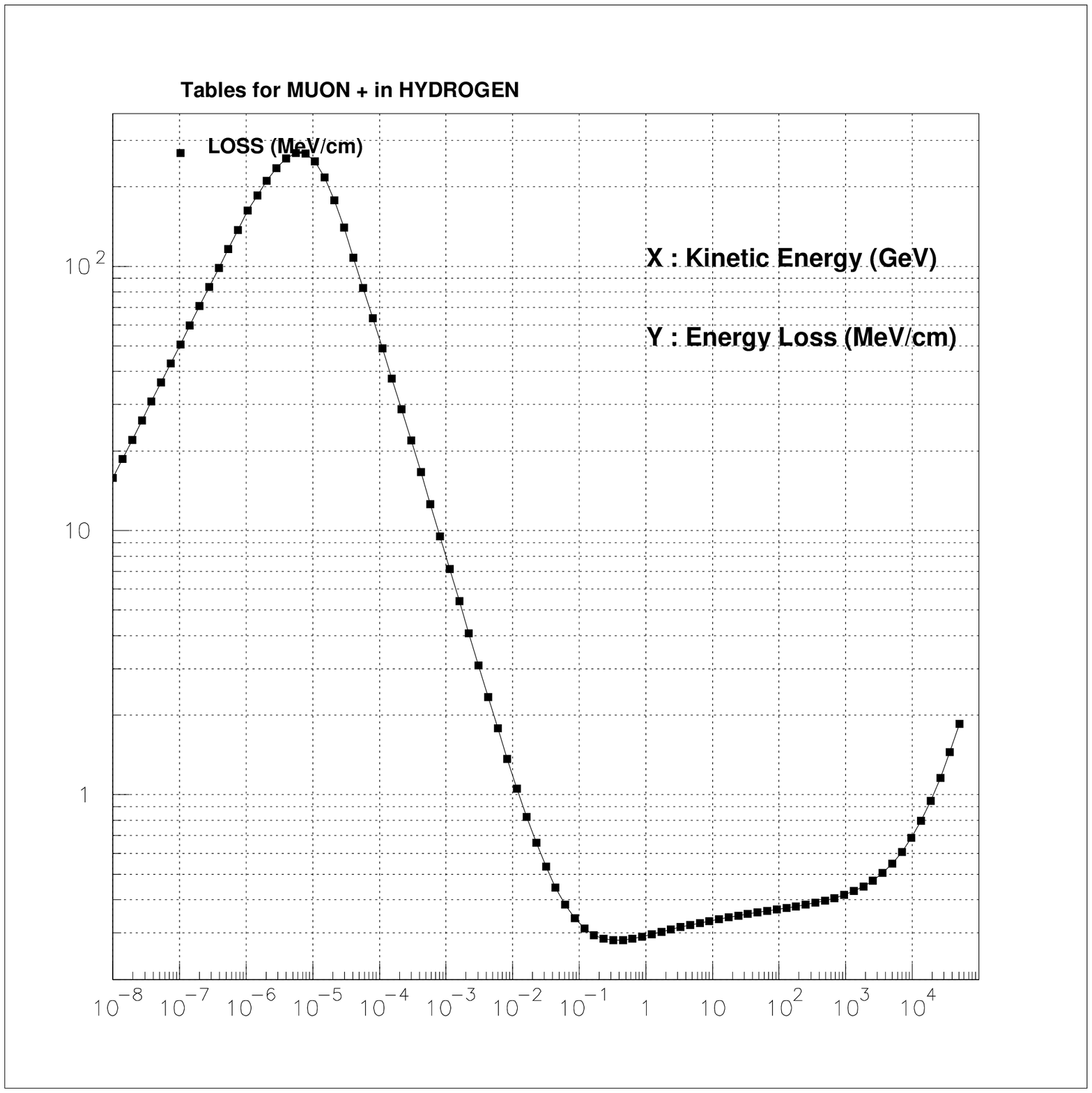,width=49mm}}
\vspace*{-10mm}
\caption{
Plot of $\mu^+$ energy loss (MeV/cm) in liquid hydrogen versus kinetic 
energy (GeV) generated
using GEANT3. The default value of the GEANT3 variable ``CUTMUO" was
decreased
from 10 MeV to 10 eV to permit the propagation of slow muons. 
The energy turnover at 8 keV corresponds to a momentum of
1.3 MeV/c. $p = \sqrt{2mE} = \sqrt{2 \times 105.7 \times 0.008}$.
Liquid helium, aluminum, copper, and iron show similar results.}
\end{minipage} \hfill
\begin{minipage}{0.58\textwidth}
\centerline{\psfig{file=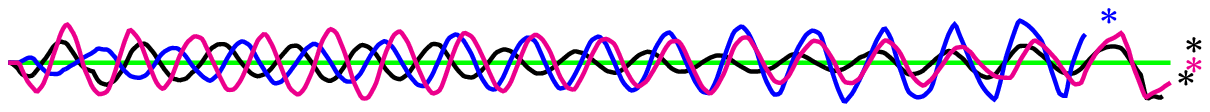,width=73mm}}
\centerline{\psfig{file=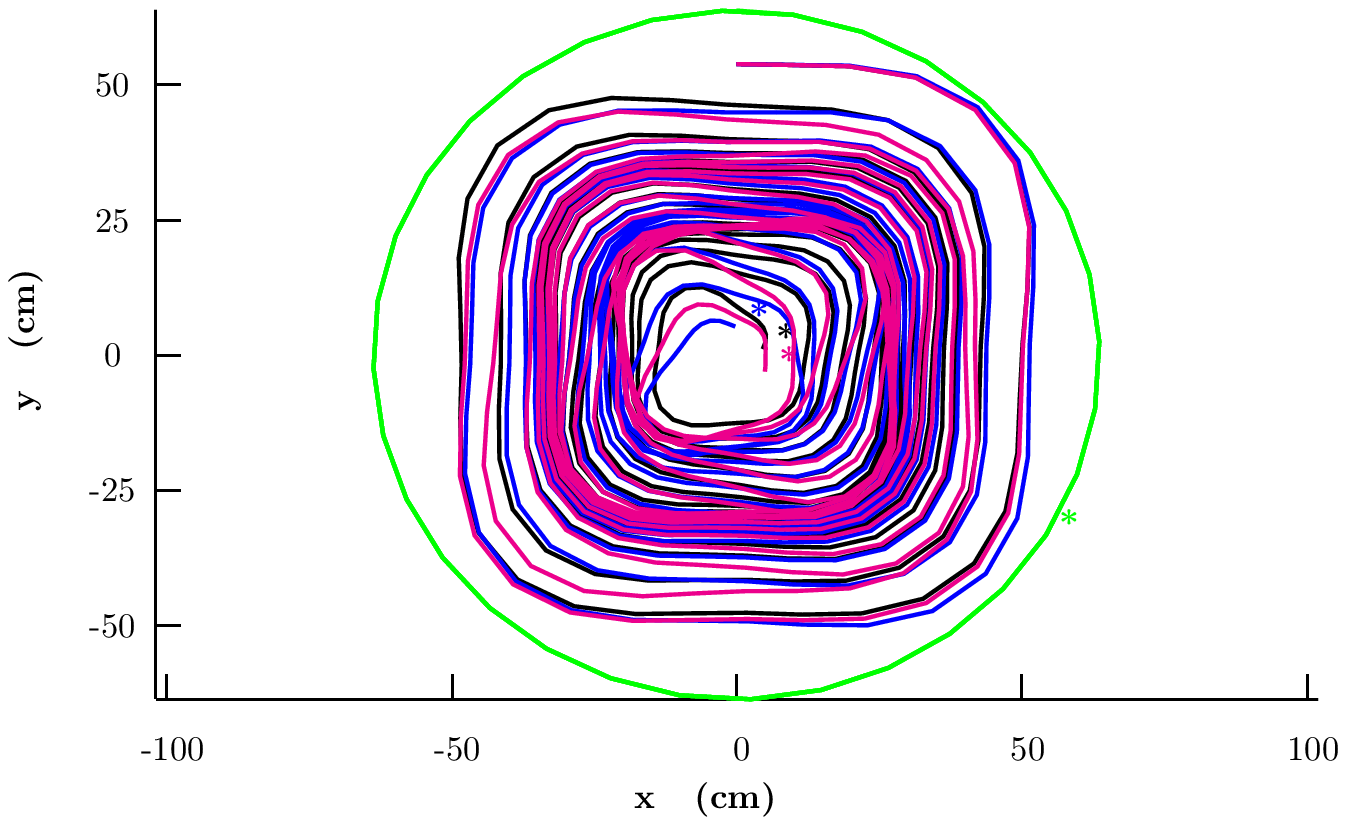,width=73mm}}
%\vspace*{8pt}
\caption{ICOOL$^{\,13}$ simulation of single turn, energy loss injection. 
Three muons with identical momenta are injected into a low field cyclotron
with four sectors like the one in
Table 1, but with soft edged magnetic fields. 
The inward spirals differ
because of multiple scattering and straggling. The energy loss is caused by
radial 
$LiH$ wedges surrounded by hydrogen gas.  The amount of matter encountered in
a given orbit decreases adiabatically with radius.
The upper trace shows that vertical motion is completely 
contained within $\pm$5\,cm along the 70\,m spiral.
The fractional energy loss required in the first turn for injection increases
with the width of the muon beam and decreases as the cyclotron's 
magnetic field is lowered.}
\end{minipage}
\end{figure}

\section{Sector Anti--Cyclotrons with External RF}
\vspace*{-2pt}

An anti-cyclotron has been used to slow LEAR anti-protons at 
CERN.$^{16, 17}$  
An annular quasipotential well, $U(r,z)$, 
is formed which ferries anti-protons towards the
center of an azimuthally symmetric cyclotron. The radius of the annulus
decreases with the decreasing angular momentum of the $\overline{p}$. 

\begin{equation}
U(r,z) = V(r,z) - (1/(2 \eta r^2))\, (L_g / M + \eta \,  r A_{\theta})^2,
\end{equation}
where $\eta = e/M$ and $L_g = L_z - e\,r\,A_{\theta}$ is a generalized
angular momentum.  The radial well deepens with decreasing radius and the
vertical well grows shallower (see Fig.~2 of Ref.~16). 
Particles must adiabatically spiral to the center.  If dE/dx is too
large, particles will not stay in the magnetic wells. 
The final $\overline{p}$ swarm has a radius of 1.5 cm, a height of
4 cm, and a kinetic energy of 2 keV. 
A long bunch train is coalesced into a single swarm, which is roughly the 
same diameter as the incoming beam.
The spiral time is 20 $\mu$s with
0.3 mbar hydrogen and about 1 $\mu$s with 10 mbar hydrogen.
Given the dependence of the cyclotron frequency on mass, 
$f = \omega / 2 \pi = q B / 2 \pi m$,
the spiral time
for a muon is nine time less than for a $\overline{p}$.
The gas pressure in the center must be low, both to allow a particle to 
spiral all the way in before
stopping, and to allow reasonable kicker voltages for extraction.
An 80 ns electric kicker 
pulse rising to 500 V/cm in 20 ns is employed.
The $\overline{p}$'s move 32 cm in 500 ns. Given that
$F = ma$, muons will go farther. 

The anti-cyclotron has now been moved from LEAR at CERN to PSI where it is
being used to slow negative muons to kinetic energies of a few keV.$^{18}$  
Three centimeter diameter beams with 30\,000 $\mu^-$/s below 50 keV
and 0.8 cm diameter beams with 1000 $\mu^-$/s in the 3 to 6 keV kinetic
energy range are output for use.
A static
electric field continuously ejects the muons.  The energy absorber and the
electrode consist of  a single 30 $\mu$g/cm$^2$ formvar foil with a 3 nm
nickel film produced by 30 minutes of sputtering.

A scaling sector anti-cyclotron could allow greater acceptance than the
azimuthally symmetric anti-cyclotron now running at PSI. For a given
$\int{\bf{B}}{\cdot}d\,{\bf\ell}$, the ratio of the fields in the hills and
valleys can be adjusted to maximize acceptance. 
Only radial and not spiral sectors have been explored so far.
The sector anti-cyclotron may
be able to function as a damped harmonic oscillator to lower the amplitude of
horizontal and vertical betatron motion as 
a bunch train of
muons spirals into a single central swarm.
With $10^{12}$ muons in a swarm, space charge is a concern.  
A conductor might be used for neutralization.
Movement of $10^{12}$ electrons in 100~ns only requires 1.6 amps of current.
Muons must spiral in fast enough to minimize decay loss, but must not stop
before reaching the central swarm. 
So, the density of the absorber must decrease smoothly with radius.
Radial $LiH$ wedges immersed in a gas or 
high to
low pressure gases separated by beam pipes might meet this criteria. The
sector cyclotron geometry must transform into an azimuthally symmetric magnetic
bottle as the muons approach the central swarm.  Otherwise, as shown by GEANT3
simulations, muons will escape though the valleys. In the transition
region the field might roughly resemble a hexapole or octupole field 
as used in an
Electron Cyclotron Resonance Ion Source (ECRIS).$^{19}$
Busch's theorem (eqn. 3)$^{\,20}$ 
has the effect of increasing the emittance as muons leave a magnetic field.
A half Tesla field and a 10 cm radius give an 8 MeV/c azimuthal kick. 
One might be able to use radial iron fins in the exit port to alleviate this
effect or reverse and increase the magnitude of the magnetic field 
to capture the unwanted angular momentum in 
an absorber just after extraction.  
Using low fields with cylindrical swarms that have small diameters 
works for sure.
An RF quadrupole is perhaps a natural choice for acceleration
that would immediately follow the extraction electric kicker.

\begin{equation}
\dot{\phi} = [e / (2 \pi \, \gamma  \, m \, r^2(s))] [\Phi(s) - \Phi_k], \quad
L_z = x p_y - y p_x = r^2 \gamma \, m \, \dot{\phi} = -e \, B \, r^2 / 2 
\end{equation}

In summary, progress on large acceptance tabletop muon rings  
is underway, including
energy loss injection (see Fig. 4), 6D cooling, and electric kicker extraction.
Many thanks to Juan Gallardo and Franz Kottmann for useful suggestions.

\begin{table}[t]
\begin{minipage}{0.30\textwidth}
\tbl{Slow muon sticking solutions. 
Charged foils might repel muons and prevent them from stopping in the foils.
In muon catalyzed fusion, a $\mu^-$ is typically freed 150 times before
it sticks to an ${^4}He$ nucleus.
}
{\begin{tabular}{@{}l@{}} \toprule
Positively charged foils for $\mu^+$. \\
Helium to slow $\mu^+ e^-$ formation. \\
Laser disassociate $\mu^+ e^-$. \\ \colrule
Negatively charged foils for $\mu^-$. \\
$DT$ or $D_2$ gas. Fusion frees $\mu^-$. \\ 
\botrule 
\end{tabular}}
\end{minipage}\hfill
\begin{minipage}{0.64\textwidth}
\tbl{Emittance reduction goals for two anti-cyclo\-trons used in series.
Between the anti-cyclotrons, Busch's theorem gives a $\Delta{p}_{\phi}$ kick,
muons are accelerated, and $\Delta p_x$ is traded for  $\Delta x$ and $\Delta
p_y$ for $\Delta y$. Emittance goes as $(\Delta p_x \, \Delta x) \, (\Delta
p_y \, \Delta y) \, (\Delta p_z \, \Delta z)$. A muon collider needs a factor
of $10^6$ in cooling.
}
{\begin{tabular}{@{}lcc@{}} \toprule
        & First Anti-Cyclotron \hspace{6mm} & Second Anti-Cyclotron \\ \colrule
${\Delta}p_x$ \hspace{4mm} & 50  $\to$ 1 MeV/c     &  15  $\to$ 1 MeV/c \\
${\Delta}p_y$       & 50  $\to$ 1 MeV/c   &  15  $\to$ 1 MeV/c \\
${\Delta}p_z$       & 50  $\to$ 1 MeV/c     &   1  $\to$ 1 MeV/c \\
${\Delta}x$       & 15    $\to$ 10 cm     &   5  $\to$ 3 cm \\
${\Delta}y$       & 15    $\to$ 10 cm     &   5  $\to$ 3 cm \\
${\Delta}z$       & 1000  $\to$ 10 cm     &  10  $\to$ 3 cm \\
\botrule 
\end{tabular}}
\end{minipage}
\end{table}

\end{document}